\documentclass[aps,pre,twocolumn,groupedaddress]{revtex4-1}
\usepackage{graphicx}
\usepackage{amsmath}
\graphicspath{{./}}

\begin{document}

\title{Multistep Bloch-line-mediated Walker breakdown in ferromagnetic strips}

\author{Johanna H\"utner$^{1,2}$}
\author{Touko Herranen$^{1}$}
\author{Lasse Laurson$^{1,3}$}
\email{lasse.laurson@tuni.fi}

\affiliation{$^{1}$Helsinki Institute of Physics and Department of Applied 
Physics, Aalto University, P.O.Box 11100, FI-00076 Aalto, Espoo, Finland}
\affiliation{$^{2}$Aalto Science Institute, Aalto University, P.O.Box 11100, 
FI-00076 Aalto, Espoo, Finland}
\affiliation{$^{3}$Computational Physics Laboratory, Tampere University, 
P.O. Box 692, FI-33014 Tampere, Finland}

\date{\today}

\begin{abstract}
A well-known feature of magnetic field 
driven dynamics of domain walls in ferromagnets is the existence of a threshold
driving force at which the internal magnetization of the domain wall starts
to precess -- a phenomenon known as the Walker breakdown -- resulting in an 
abrupt drop of the domain wall propagation velocity. Here, we report on 
micromagnetic simulations of magnetic field driven domain wall 
dynamics in thin ferromagnetic strips with perpendicular magnetic anisotropy 
which demonstrate that in wide enough strips Walker breakdown is a 
multistep process: It consists of several distinct velocity drops separated
by short linear parts of the velocity vs field curve. These features 
originate from the repeated nucleation, propagation and annihilation of an 
increasing number of Bloch lines within 
the domain wall as the driving field magnitude is increased. This mechanism 
arises due to magnetostatic effects breaking the symmetry between the 
two ends of the domain wall.  
\end{abstract}


\maketitle

\section{Introduction}

Domain wall (DW) dynamics driven by applied magnetic fields 
\cite{beach2005dynamics,metaxas2007creep,schryer1974motion}
or spin-polarized
electric currents \cite{parkin2008magnetic,thiaville2005micromagnetic,
moore2008high} is an active field of research catalyzed by both 
fundamental physics interests as well as promising applications
in technology. One of the most striking features of DW dynamics
is that one typically observes a non-monotonic driving force dependence
of the DW propagation velocity $v_\text{DW}$. Considering 
field-driven DW dynamics, for small applied
fields $B_\text{ext}$, $v_\text{DW}$ first increases with $B_\text{ext}$, 
followed by a sudden drop of $v_\text{DW}$. The latter originates from an 
instability known as the Walker breakdown \cite{schryer1974motion}, where the 
internal DW magnetization starts precessing at $B_\text{ext}=B_\text{W}$,
with $B_\text{W}$ known as the Walker field. This leads to a reduced $v_\text{DW}$ 
for $B_\text{ext} > B_\text{W}$ as part of the energy of the driving field 
is dissipated by the precessional magnetization dynamics within the DW. 

The widely used one-dimensional (1$d$) models \cite{thiaville2006domain} 
describe this precession by a single angular variable, and have been 
demonstrated to successfully capture the DW dynamics in nanowire 
geometries \cite{mougin2007domain}. However, this simple 
description fails in wide enough strips. In such systems 
an instability analogous to the Walker breakdown in nanowires is known 
to proceed in a spatially non-uniform fashion via repeated nucleation 
and propagation of Bloch lines (BLs) within the DW
\cite{herranen2015domain,herranen2017bloch,thiaville2018topology}. 
BLs are topologically stable magnetization textures 
corresponding to localized transition regions separating different
chiralities of the Bloch DW. In the case of thin strips considered 
here, BLs are lines threading the strip in the thickness direction, 
and are hence referred to as vertical Bloch lines 
(VBLs) \cite{thiaville2018topology,garanin2017skyrmion}.
Even if the study of BLs especially in the context of bubble
materials has a long history dating back to the 1970's 
\cite{malozemoff2016magnetic,konishi1983new}, the various BL excitation 
modes responsible for the velocity drop in strips of different geometries 
remain to be understood.

\begin{figure}[ht!]
\includegraphics[width=0.8\columnwidth]{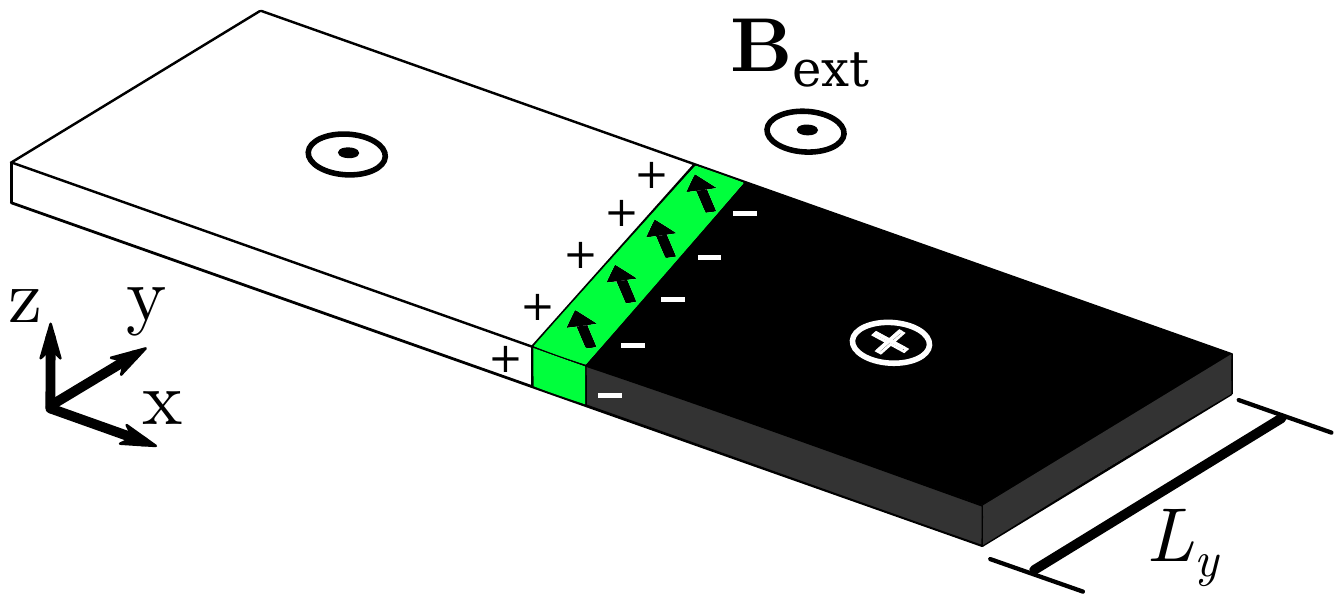}
\caption{Schematic representation of the simulated system.
Two out-of-plane polarized domains are separated by a DW,
which in equilibrium is a pure Bloch wall. As illustrated in 
the figure, upon application of
an out-of-plane magnetic field $B_\text{ext}<B_\text{W}$, the
magnetization of the moving DW finds a steady state orientation
corresponding to a partial N\'eel wall structure (arrows), producing 
magnetic charges on the DW surfaces. To minimize the resulting
magnetostatic energy, the DW tries to orient itself with the
DW magnetization, leading to DW tilting.}
\label{fig:1}
\end{figure}

Hence, we perform here extensive micromagnetic simulations of 
field-driven DW dynamics considering thin CoPtCr strips with strong 
perpendicular magnetic anisotropy 
as example systems (see Fig. \ref{fig:1}). We study in detail the dependence of 
the DW propagation velocity $v_\text{DW}$ on the applied field $B_\text{ext}$,
as well as the onset of precessional dynamics at 
$B_\text{ext} = B_\text{W}$ for a wide range of strip 
widths $L_y$. Remarkably, by carefully inspecting the ``fine structure'' of the
Walker breakdown, we find that for wide enough strips the large
velocity drop in the $v_\text{DW}(B_\text{ext})$ curve observed 
previously \cite{herranen2015domain} actually consists of several 
distinct, smaller velocity drops, separated by short linearly increasing 
parts of $v_\text{DW}(B_\text{ext})$. Our analysis of the corresponding 
VBL dynamics within the DW shows that this behaviour arises due to a 
sequence of distinct excitations of the DW magnetization. Thereby, 
the number of VBLs present within the DW increases with $B_\text{ext}$ in 
discrete steps at specific $B_\text{ext}$-values. We show that these 
features are a consequence of DW tilting due to magnetostatic 
effects, breaking the symmetry between the two ends of the DW.

The paper is organized as follows: In Sec. \ref{sec:1} we go through
the details of our micromagnetic simulations, while in Sec. \ref{sec:2}
we present our results, focusing on the multistep nature of the 
Walker breakdown in wide strips. Sec. \ref{sec:3} finishes the paper
with conclusions.

\section{Simulations}
\label{sec:1}

Our micromagnetic simulations are performed using the GPU-accelerated
micromagnetic simulation program MuMax3 \cite{vansteenkiste2014design}. 
It solves the space and 
time-dependent reduced magnetization ${\bf m}({\bf r},t)=
{\bf M}({\bf r},t)/M_\text{s}$ [with ${\bf M}({\bf r},t)$ and 
$M_\text{s}$ the magnetization and saturation magnetization, 
respectively] from the Landau-Lifshitz-Gilbert (LLG) equation, 
\begin{equation}
\label{eq:1}
\frac{\partial {\bf m}}{\partial t} = -\frac{\gamma}{1+\alpha^2}
\left[{\bf m}\times {\bf B}_\text{eff} + \alpha ({\bf m} \times 
({\bf m} \times {\bf B}_\text{eff}))\right],
\end{equation}
using a finite-difference discretization. In Eq. (\ref{eq:1}), 
$\gamma$ is the gyromagnetic ratio, $\alpha$ the dimensionless
damping parameter and ${\bf B}_\text{eff}$ the effective field
having contributions from the externally applied field 
${\bf B}_\text{ext}$, magnetostatic field, Heisenberg exchange
field as well as the anisotropy field.
As a test system, we consider CoPtCr strips of thickness
$L_z = 12$ nm and widths $L_y$ ranging from 90 nm to 1800 nm.
The length of the moving simulation window centered around the 
DW (implying that the dipolar fields due to the two domains cancel
at the domain wall) is $L_x = 3072$ nm. The system is discretized 
using cubic discretization cells with a side length of 3 nm.
The typical material parameters of CoPtCr \cite{weller2000high,
herranen2015domain} used here are uniaxial magnetic
anisotropy $K_\mathrm{u} = 2 \times 10^5 \, \mathrm{J/m^3}$,
exchange constant $A_\mathrm{ex} = 10^{-11}$ J/m, damping parameter 
$\alpha = 0.2$, and saturation magnetization $M_\mathrm{s} = 3 
\times 10^5 $ A/m, corresponding to the stray field energy constant
of $K_\text{d} = \mu_0 M_\text{s}^2/2 = 5.65 \times 10^4 \, \mathrm{J/m^3}$, 
where $\mu_0$ is the vacuum permeability.
These values result in the Bloch 
wall width parameter $\Delta = \sqrt{A_\mathrm{ex}/K_\text{u}} 
\approx 7.1$ nm and the Bloch line width parameter (or the exchange length) 
$\Lambda = \sqrt{A_\mathrm{ex}/K_\text{d}} \approx 13.3$ nm.

\begin{figure}[ht!]
\includegraphics[width=\columnwidth]{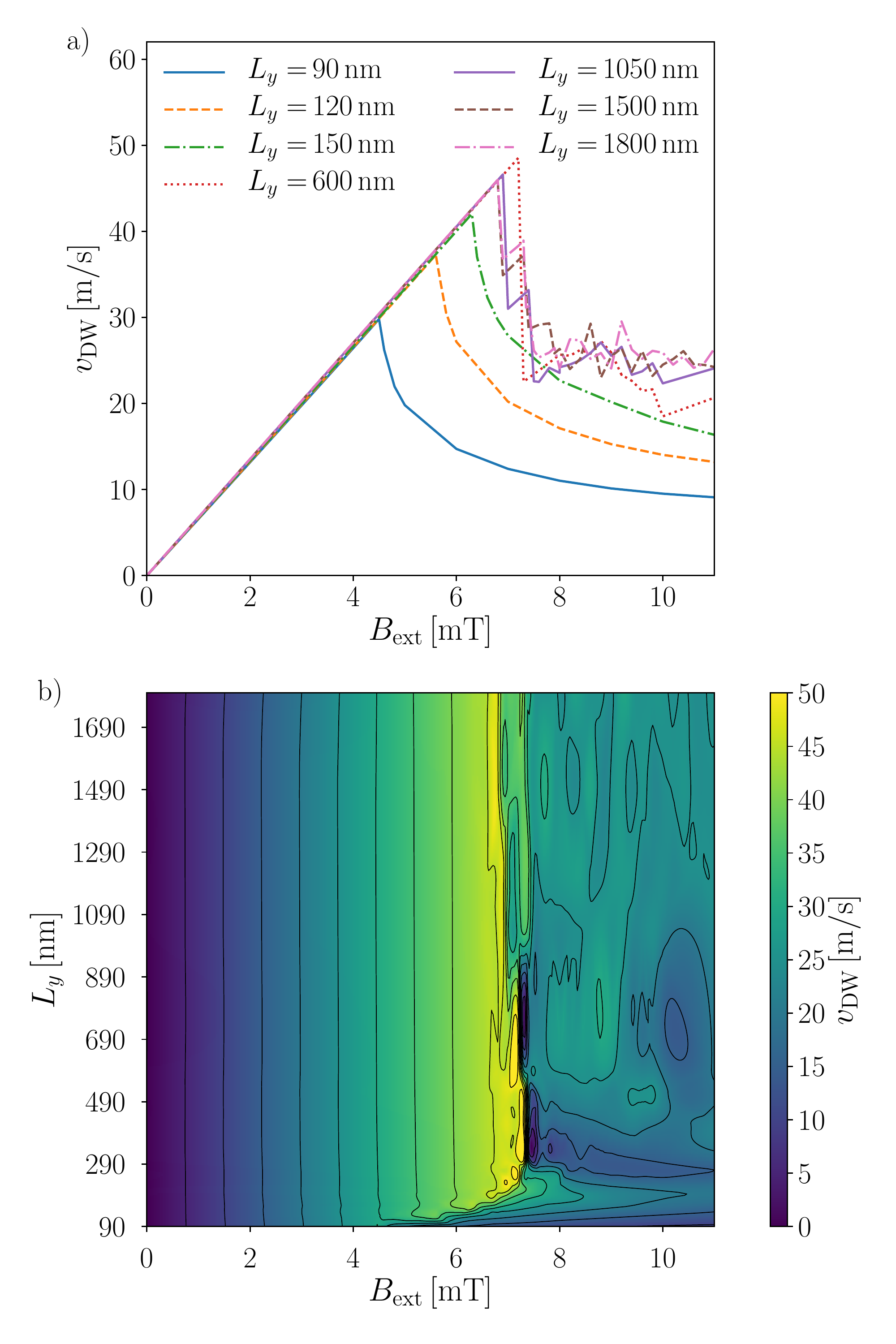}
\caption{a) $v_\mathrm{DW}$ as a function
of $B_\mathrm{ext}$ considering a representative subset of different $L_y$'s. 
Note the ``smooth'' velocity drop for narrow strips that changes first to a 
single large drop ($L_y=600$ nm) and then develops two or even three distinct 
velocity drops separated by short linear parts of the $v_\text{DW}(B_\text{ext})$
curve upon increasing $L_y$. b) All the simulated $v_\mathrm{DW} (B_\mathrm{ext})$ 
data visualized as a contour plot, highlighting the non-monotonic dependence 
of $B_\text{W}$ on $L_y$.}
\label{fig:2}
\end{figure}

The system is initialized in a configuration with two antiparallel
out-of-plane ($\pm z$) domains separated by a straight Bloch DW 
with the DW internal magnetization in the positive $y$-direction. 
The DW spans the strip width along the 
$y$-direction and is located in the middle of the sample.
Upon sharp application of an external magnetic field $B_\text{ext}$ along
the positive $z$-direction, the DW is displaced in the positive $x$-direction.
The steady state time-averaged DW velocities are then estimated from the slopes
of the DW position vs time graphs, averaging over several cycles
of the precessional DW dynamics for $B_\text{ext}>B_\text{W}$ and excluding
any initial transients. 

At this point we note a crucial feature of field-driven DW dynamics
in the strip geometry, illustrated in Fig. \ref{fig:1}: A $B_\text{ext}$ 
smaller than the Walker field $B_\text{W}$ tends to rotate the DW 
magnetization counterclockwise away from the positive $y$-direction (i.e.,
away from a pure Bloch wall configuration), such 
that the moving steady state DW acquires a N\'eel component (a finite
$x$-component of the DW magnetization). This 
results in magnetic charges on the DW surfaces, with an associated 
cost in demagnetization energy. To minimize this energy, the DW tends
to tilt in an attempt to align itself with the DW magnetization.
A balance between the DW energy (proportional to the DW length) and
the magnetostatic energy leads to a finite steady state DW tilt angle
(see Fig. \ref{fig:1}).
This mechanism will be crucial for understanding the properties
of the Walker breakdown in the case of wide strips, discussed later 
in this paper.

\section{Results}
\label{sec:2}

We start by considering the relation between DW propagation velocity
$v_\text{DW}$ and $B_\text{ext}$ for strips of different widths.
Fig. \ref{fig:2}a shows examples of $v_\text{DW} (B_\text{ext})$ curves,
illustrating the key aspects of the observed DW dynamics. For all
strip widths the usual linear dependence of $v_\text{DW}$ on 
$B_\text{ext}$ for small $B_\text{ext}$ is terminated at an 
$L_y$-dependent Walker field $B_\text{W}$. This is also depicted
in the contour plot shown in Fig. \ref{fig:2}b. 
$B_\text{W}$ first increases rapidly with $L_y$, reaches a maximum for 
$L_y \approx 350$ nm, after which $B_\text{W}$ slowly decreases, 
possibly reaching a plateau for the largest $L_y$-values considered. 
This non-monotonic $L_y$-dependence is reminiscent of our recent
results on thickness-dependent Walker breakdown in garnet strips 
\cite{herranen2017bloch}, and will be analyzed further below.

\begin{figure}[t!]
\includegraphics[width=\columnwidth]{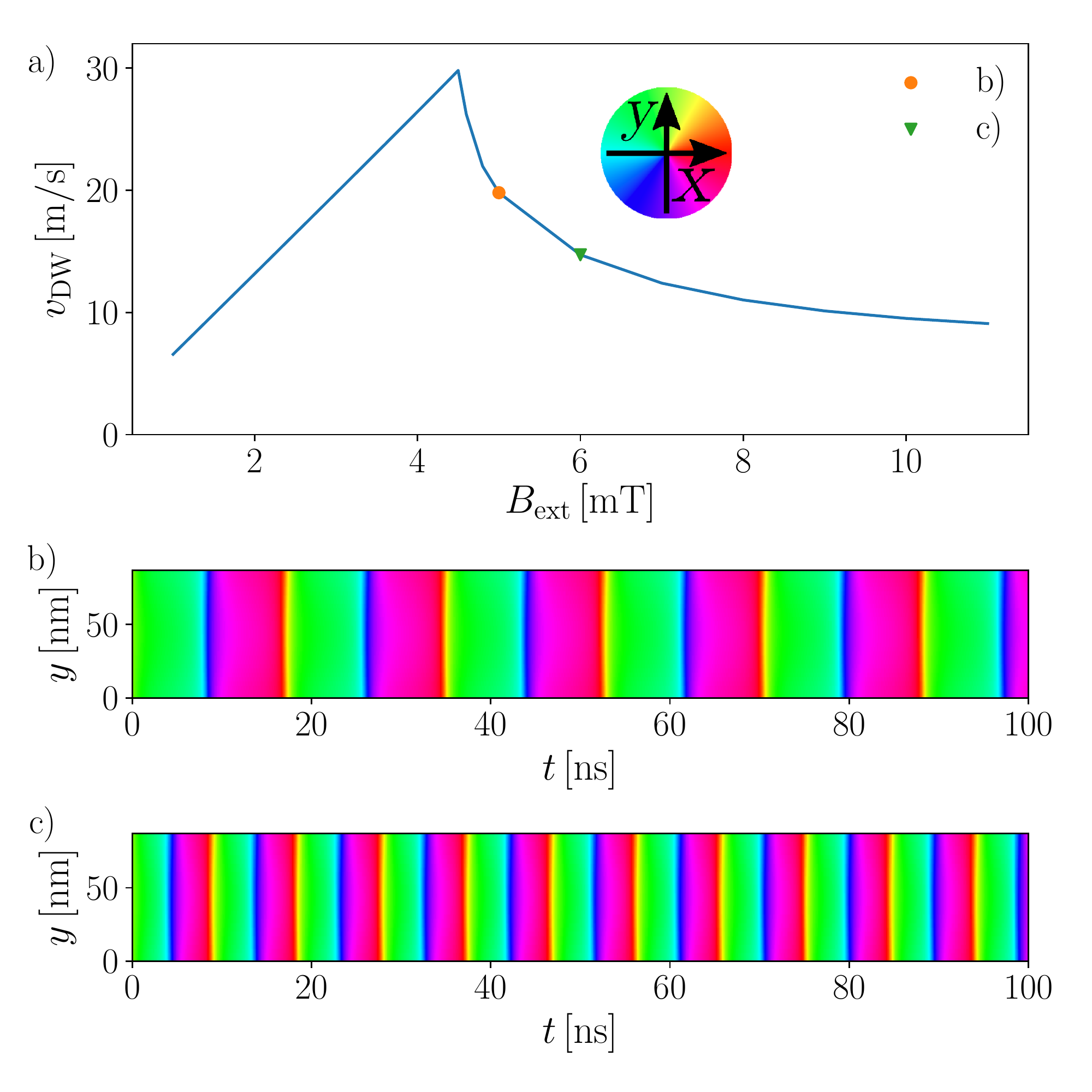}
\caption{a) An example of a typical $v_\text{DW}(B_\text{ext})$
curve of narrow strips (here, the $L_y$ = 90 nm case is shown),
exhibiting a ``smooth'' velocity drop for $B_\text{ext}>B_\text{W}$.
b) and c) display space-time maps of the internal DW magnetization
(with the colorwheel indicating the mapping from colors to magnetization)
for two $B_\text{ext}$ values (5 and 6 mT, respectively) as indicated
in a) with the two symbols. These describe the time-evolution of the
internal in-plane magnetization of the domain wall for different
$y$-coordinates along the domain wall. 
The in-plane DW magnetization exhibits
coherent (spatially uniform) periodic switching events, with the
frequency of the events increasing with $B_\text{ext}$.}
\label{fig:3}
\end{figure}

The shape of the $v_\text{DW} (B_\text{ext})$ curve
displaying the velocity drop crucially depends on $L_y$. For small
$L_y$, corresponding to the regime where $B_\text{W}(L_y)$
increases with $L_y$ (Fig \ref{fig:2}b), $v_\text{DW}$ decreases 
smoothly and gradually 
with increasing $B_\text{ext}$ (Figs. \ref{fig:2}a and \ref{fig:3}a). 
Figs. \ref{fig:3}b and \ref{fig:3}c display space-time maps of the
DW internal in-plane magnetization during the dynamics; for each $y$-coordinate
along the DW the magnetization shown is that of the mid-point of the DW where 
$m_z$ changes sign when moving along the $x$-direction. 
These maps show that above $B_\text{W}$ the internal dynamics
within the DW display the typical periodic switching of 
the DW magnetization \cite{martinez2011stochastic}, 
with the frequency of the switching events increasing with $B_\text{ext}$.
Notably, for the rather narrow system  with $L_y = 90$ nm (i.e., not 
much wider than the BL width $\pi\Lambda \approx$ 42 nm) studied
in Fig. \ref{fig:3}, these switching events are to a very good
approximation spatially uniform, such that the magnetization of the 
entire DW rotates synchronously, and no VBLs are observed.

\begin{figure*}[ht!]
\includegraphics[width=\textwidth]{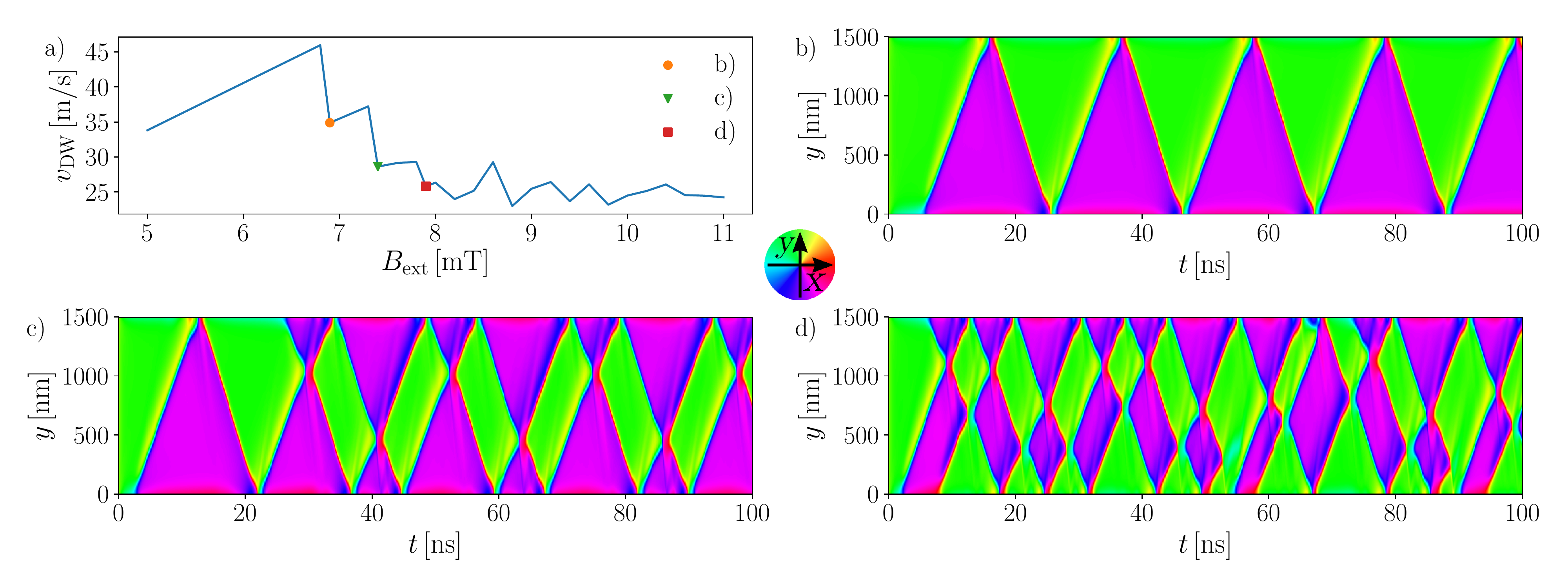}
\caption{A closer look at the DW dynamics corresponding to
a multistep Walker breakdown for a strip of width $L_y = 1500$ nm.
The $v_\text{DW} (B_\text{ext})$ curve shown in a) exhibits
three distinct velocity drops, with the corresponding VBL
dynamics illustrated by means of space-time maps of the DW internal
in-plane magnetization (with the colorwheel in the middle showing the
mapping from colors to magnetization direction) in b), c) and d). 
b) shows a single VBL first nucleating from the bottom edge (i.e.,
the leading end of the DW just before the onset of Walker breakdown),
and then travelling back and forth along the DW. These dynamics are
responsible for the first velocity drop seen in a). c) displays
the VBL dynamics corresponding to the second velocity drop,
where after an initial transient a VBL is first nucleated from
the bottom edge, and shortly afterwards a second VBL is
nucleated from the top edge. Their annihilation is followed by an 
almost immediate formation of another pair of VBLs that
propagate to the edges, after which the process repeats. 
d) shows the dynamics corresponding to the third velocity drop, 
involving the simultaneous presence of three VBLs within the DW.
Movies illustrating the dynamics shown in b), c) and d) are included
as Supplemental Material \cite{SM}.}
\label{fig:4}
\end{figure*}

This is in strong contrast to the behaviour
in wider strips [$L_y$ $\gg \pi\Lambda$ and beyond the 
maximum of $B_\text{W}(L_y)$]:
First, when increasing $L_y$, a single, quite steep velocity drop 
is observed; an example is given by the $L_y = 600$ nm curve in Fig. 
\ref{fig:2}a. For even wider strips, a remarkable feature is observed:
Our simulations where we consider a finer sampling of the $B_\text{ext}$ 
values than previous studies \cite{herranen2015domain} reveal that the 
Walker breakdown actually consists of multiple distinct velocity drops, 
separated by short linear parts of the $v_\text{DW} (B_\text{ext})$ curve.
First, for $L_y = 1050$ nm (Fig. \ref{fig:2}a), we observe two
velocity drops, and further increasing $L_y$ to 1500 nm leads to
the appearence of three of these steps. All velocity
drops take place within a rather narrow field range of less than
1 mT (they all occur between 6.9 and 7.9 mT). Thus, they were not
clearly observed in previous work \cite{herranen2015domain},
where the sampling of the $B_\text{ext}$-values was much more 
coarse.

To account for these distinct velocity drops, it is again
instructive to consider the details of the underlying DW 
magnetization dynamics. Fig. \ref{fig:4}a shows an example
of a $v_\text{DW} (B_\text{ext})$ curve exhibiting three velocity
drops, followed by a more irregular structure for larger $B_\text{ext}$
($L_y = 1500$ nm). Subsequent to the first velocity drop 
(for $B_\text{ext} \approx$ 6.9 mT), 
as illustrated in the space-time map of DW internal magnetization in 
Fig. \ref{fig:4}b, a single VBL 
nucleates from the bottom edge of the strip, propagates along the DW 
across the strip width, exits the strip, after which another VBL 
of opposite $x$-magnetization (shown in red instead of blue in 
Fig. \ref{fig:4}b) enters the strip/DW and propagates to the 
opposite strip edge, before the process repeats.
Upon increasing $B_\text{ext}$ to $B_\text{ext} \approx$ 
7.4 mT, a second velocity
drop occurs. Fig. \ref{fig:4}c shows that this second drop
is due to more complex VBL dynamics within the DW: After an
initial transient, the system finds a steady state where
another VBL is nucleated from the top strip edge before 
the VBL nucleated from the bottom edge reaches the top edge. 
Subsequently, the two VBLs annihilate within the strip, 
and a new pair of VBLs is created in the same DW segment. 
These two VBLs then 
propagate towards the bottom and top edges of the strip,
respectively, and exit the strip. Thereafter, the process is repeated.
A third velocity drop is observed for $B_\text{ext} \approx$ 7.9 mT,
with the corresponding DW magnetization dynamics 
shown in Fig. \ref{fig:4}d: In this case, three VBLs are present 
within the DW for most of the time. Upon further increasing
$B_\text{ext}$, the VBL dynamics become increasingly
complex (not shown) and no further clear, distinct velocity drops 
can be resolved (Fig. \ref{fig:4}a). Movies illustrating 
the DW dynamics shown in Fig. \ref{fig:4}b, c and d are
included as Supplemental Material \cite{SM}. Notice that 
while Figs. \ref{fig:4} b, c and d describe the VBL dynamics along 
the DWs, the movies show in addition that DWs containing VBLs 
are not straight lines but tend to exhibit significant curvature
especially at the locations of the VBLs.

The described dynamics of VBLs responsible for the 
distinct velocity drops crucially depend on a broken symmetry
between the two ends of the DW (bottom vs top strip edges). As
illustrated in Fig. \ref{fig:1}, for $B_\text{ext} < B_\text{W}$,
the driving field rotates the magnetization of the moving DW away 
from a pure Bloch wall configuration 
to a steady DW structure with a finite N\'eel component. 
The N\'eel nature of the
DW gives rise to magnetic charges at the DW surfaces (Fig. \ref{fig:1}). 
To reduce the resulting energy, the DW develops a tilt as it
attempts to minimize the charges by aligning with its internal 
magnetization. Thus, the leading end of the DW effectively experiences 
a larger driving force (sum of $B_\text{ext}$ and the 
demagnetizing fields due to the DW surface charges) than the trailing 
one. Hence, when incresing $B_\text{ext}$ over the Walker threshold,
the leading end of the DW experiences the breakdown first, i.e., 
at a lower $B_\text{ext}$, while the trailing end is still below 
its (local) Walker breakdown field.
This means that the first VBL is always nucleated from the
leading end of the DW (bottom edge in Figs. \ref{fig:4}b-d), 
and that the first velocity drop corresponds to a single VBL
moving back and forth along the DW (Fig. \ref{fig:4}b).

When $B_\text{ext}$ is increased to reach the 
second velocity drop, also the trailing end of the DW exceeds its
local Walker threshold, and VBLs are nucleated from both
ends of the DW. The leading end of the DW still 
experiences a larger effective driving force, and hence, the first
VBL is nucleated from this edge. However, before it reaches
the other end of the DW, a second VBL is nucleated from the
trailing end, and subsequently the two Bloch lines
annihilate inside the strip, followed by creation of a new pair 
of VBLs in the same location (Fig. \ref{fig:4}c). Increasing
$B_\text{ext}$ even more to reach the third velocity drop leads
to nucleation of a third VBL, while the two first ones are
still inside the strip, resulting in the simultaneous presence of
three VBLs along the DW (Fig. \ref{fig:4}d). We note that
all creation and annihilation reactions in Fig. \ref{fig:4} 
respect the conservation of the magnetic charge $Q = \pm 1$ and 
chirality $C = \pm 1/2$ of the four-fold degenerate VBLs 
\cite{yoshimura2016soliton}.

Finally, we address the non-monotonic dependence of the Walker 
field $B_\text{W}$ (defined as the $B_\text{ext}$ value where the
first velocity drop takes place) on $L_y$ (see Fig. \ref{fig:2}). 
As found by Mougin {\it et al.}
\cite{mougin2007domain}, in confined geometries with uniform magnetization
along the DW $B_\text{W} \propto
|N_x-N_y|$, where $N_x$ and $N_y$ denote the demagnetizing factors
of the DW along $x$ and $y$, respectively. Employing the elliptic approximation
leads to $N_x \approx L_z/(L_z + \pi \Delta)$ 
and $N_y \approx L_z/(L_z + L_y)$ \cite{mougin2007domain,boulle2011current}. 
Notice that the DW width $\pi \Delta \approx 22.3$ nm used above can be obtained 
by integrating the Bloch wall profile $m_y = 1/\cosh (x/\Delta)$ 
\cite{hubert2008magnetic}, and the approximate expressions for $N_x$ and 
$N_y$ utilized are valid for $L_z \ll \pi \Delta$ and $L_z \ll L_y$, 
respectively. Thus, we obtain the approximate result that 
$B_\text{W} \propto |L_z/(L_z + \pi \Delta) - L_z/(L_z + L_y)|$, suggesting
that $B_\text{W}$ increases with $L_y$, in agreement with our 
observations for narrow strips (small $L_y$), where the 
magnetization of the entire DW precesses in phase above the breakdown 
(see Fig. \ref{fig:3}). However, the above expression also predicts a
saturation of $B_\text{W}$ in the limit $L_y \gg L_z$, at
odds with our observation in Fig. \ref{fig:2} where, after reaching a
maximum, $B_\text{W}$ is slowly decreasing with $L_y$. Indeed, 
the calculation in \cite{mougin2007domain} is valid for uniform DW magnetization 
only. In particular, it does not take into account the possibility of 
nucleation of VBLs within the DW which is the mechanism underlying the 
Walker breakdown for large $L_y$. The energy barrier for VBL 
nucleation should depend on $L_y$, such that it is lower for longer 
DWs (larger $L_y$). However, for the very largest strip widths $L_y$ considered 
(1500 and 1800 nm), $B_\text{W}$ appears to saturate to a value of 
$B_\text{W} \approx 6.7$ mT.   

\section{Conclusions}
\label{sec:3}

Thus, we have established that precessional DW dynamics in 
PMA strips undergo a transition from spatially homogeneous
precession of the DW magnetization to a VBL-dominated
regime as the strip width $L_y$ is increased. The latter regime
is characterized by multiple distinct velocity drops in the
$v_\text{DW}(B_\text{ext})$ curve, originating from asymmetric nucleation
of VBLs from the strip edges due to DW tilting. This closer look at 
the well-studied phenomenon of Walker breakdown thus reveals its 
multistep nature for DWs with lengths well above the VBL width. 
These features should lend themselves to experimental verification in 
future studies. It would also be of interest to extend our study to 
systems with structural disorder or inhomogeneities interacting with
the DW \cite{leliaert2014numerical,leliaert2014current}, to consider
the possible effects of a small tilt of the applied field, as well
as to investigate other materials characterized by different
micromagnetic parameters; considering such details numerically would
be helpful in better understanding the experimental conditions
where the mechanism reported here could be observed. We would expect
that the multi-step nature of Walker breakdown should be experimentally
observable whenever the disorder-induced depinning field is well below
the Walker field. Another 
future avenue of research of considerable current interest would be to 
address the effect of a finite Dzyaloshinskii-Moriya interaction (DMI) 
\cite{thiaville2012dynamics}, resulting in a scenario where the 
degeneracy of the different VBL configurations is lifted due to 
DMI-induced splitting of the energy levels \cite{yoshimura2016soliton}.
 
\begin{acknowledgments}
This work has been supported by the Academy of Finland through 
an Academy Research Fellowship (LL, project no. 268302). We acknowledge the 
computational resources provided by the Aalto University School of Science 
``Science-IT'' project, as well as those provided by CSC (Finland).
\end{acknowledgments}

\bibliography{refs}

\begin{thebibliography}{24}%
\makeatletter
\providecommand \@ifxundefined [1]{%
 \@ifx{#1\undefined}
}%
\providecommand \@ifnum [1]{%
 \ifnum #1\expandafter \@firstoftwo
 \else \expandafter \@secondoftwo
 \fi
}%
\providecommand \@ifx [1]{%
 \ifx #1\expandafter \@firstoftwo
 \else \expandafter \@secondoftwo
 \fi
}%
\providecommand \natexlab [1]{#1}%
\providecommand \enquote  [1]{``#1''}%
\providecommand \bibnamefont  [1]{#1}%
\providecommand \bibfnamefont [1]{#1}%
\providecommand \citenamefont [1]{#1}%
\providecommand \href@noop [0]{\@secondoftwo}%
\providecommand \href [0]{\begingroup \@sanitize@url \@href}%
\providecommand \@href[1]{\@@startlink{#1}\@@href}%
\providecommand \@@href[1]{\endgroup#1\@@endlink}%
\providecommand \@sanitize@url [0]{\catcode `\\12\catcode `\$12\catcode
  `\&12\catcode `\#12\catcode `\^12\catcode `\_12\catcode `\%12\relax}%
\providecommand \@@startlink[1]{}%
\providecommand \@@endlink[0]{}%
\providecommand \url  [0]{\begingroup\@sanitize@url \@url }%
\providecommand \@url [1]{\endgroup\@href {#1}{\urlprefix }}%
\providecommand \urlprefix  [0]{URL }%
\providecommand \Eprint [0]{\href }%
\providecommand \doibase [0]{http://dx.doi.org/}%
\providecommand \selectlanguage [0]{\@gobble}%
\providecommand \bibinfo  [0]{\@secondoftwo}%
\providecommand \bibfield  [0]{\@secondoftwo}%
\providecommand \translation [1]{[#1]}%
\providecommand \BibitemOpen [0]{}%
\providecommand \bibitemStop [0]{}%
\providecommand \bibitemNoStop [0]{.\EOS\space}%
\providecommand \EOS [0]{\spacefactor3000\relax}%
\providecommand \BibitemShut  [1]{\csname bibitem#1\endcsname}%
\let\auto@bib@innerbib\@empty
\bibitem [{\citenamefont {Beach}\ \emph {et~al.}(2005)\citenamefont {Beach},
  \citenamefont {Nistor}, \citenamefont {Knutson}, \citenamefont {Tsoi},\ and\
  \citenamefont {Erskine}}]{beach2005dynamics}%
  \BibitemOpen
  \bibfield  {author} {\bibinfo {author} {\bibfnamefont {G.~S.}\ \bibnamefont
  {Beach}}, \bibinfo {author} {\bibfnamefont {C.}~\bibnamefont {Nistor}},
  \bibinfo {author} {\bibfnamefont {C.}~\bibnamefont {Knutson}}, \bibinfo
  {author} {\bibfnamefont {M.}~\bibnamefont {Tsoi}}, \ and\ \bibinfo {author}
  {\bibfnamefont {J.~L.}\ \bibnamefont {Erskine}},\ }\href@noop {} {\bibfield
  {journal} {\bibinfo  {journal} {Nat. Mater.}\ }\textbf {\bibinfo {volume}
  {4}},\ \bibinfo {pages} {741} (\bibinfo {year} {2005})}\BibitemShut {NoStop}%
\bibitem [{\citenamefont {Metaxas}\ \emph {et~al.}(2007)\citenamefont
  {Metaxas}, \citenamefont {Jamet}, \citenamefont {Mougin}, \citenamefont
  {Cormier}, \citenamefont {Ferr{\'e}}, \citenamefont {Baltz}, \citenamefont
  {Rodmacq}, \citenamefont {Dieny},\ and\ \citenamefont
  {Stamps}}]{metaxas2007creep}%
  \BibitemOpen
  \bibfield  {author} {\bibinfo {author} {\bibfnamefont {P.}~\bibnamefont
  {Metaxas}}, \bibinfo {author} {\bibfnamefont {J.}~\bibnamefont {Jamet}},
  \bibinfo {author} {\bibfnamefont {A.}~\bibnamefont {Mougin}}, \bibinfo
  {author} {\bibfnamefont {M.}~\bibnamefont {Cormier}}, \bibinfo {author}
  {\bibfnamefont {J.}~\bibnamefont {Ferr{\'e}}}, \bibinfo {author}
  {\bibfnamefont {V.}~\bibnamefont {Baltz}}, \bibinfo {author} {\bibfnamefont
  {B.}~\bibnamefont {Rodmacq}}, \bibinfo {author} {\bibfnamefont
  {B.}~\bibnamefont {Dieny}}, \ and\ \bibinfo {author} {\bibfnamefont
  {R.}~\bibnamefont {Stamps}},\ }\href@noop {} {\bibfield  {journal} {\bibinfo
  {journal} {Phys. Rev. Lett.}\ }\textbf {\bibinfo {volume} {99}},\ \bibinfo
  {pages} {217208} (\bibinfo {year} {2007})}\BibitemShut {NoStop}%
\bibitem [{\citenamefont {Schryer}\ and\ \citenamefont
  {Walker}(1974)}]{schryer1974motion}%
  \BibitemOpen
  \bibfield  {author} {\bibinfo {author} {\bibfnamefont {N.~L.}\ \bibnamefont
  {Schryer}}\ and\ \bibinfo {author} {\bibfnamefont {L.~R.}\ \bibnamefont
  {Walker}},\ }\href@noop {} {\bibfield  {journal} {\bibinfo  {journal} {J.
  Appl. Phys.}\ }\textbf {\bibinfo {volume} {45}},\ \bibinfo {pages} {5406}
  (\bibinfo {year} {1974})}\BibitemShut {NoStop}%
\bibitem [{\citenamefont {Parkin}\ \emph {et~al.}(2008)\citenamefont {Parkin},
  \citenamefont {Hayashi},\ and\ \citenamefont {Thomas}}]{parkin2008magnetic}%
  \BibitemOpen
  \bibfield  {author} {\bibinfo {author} {\bibfnamefont {S.~S.}\ \bibnamefont
  {Parkin}}, \bibinfo {author} {\bibfnamefont {M.}~\bibnamefont {Hayashi}}, \
  and\ \bibinfo {author} {\bibfnamefont {L.}~\bibnamefont {Thomas}},\
  }\href@noop {} {\bibfield  {journal} {\bibinfo  {journal} {Science}\ }\textbf
  {\bibinfo {volume} {320}},\ \bibinfo {pages} {190} (\bibinfo {year}
  {2008})}\BibitemShut {NoStop}%
\bibitem [{\citenamefont {Thiaville}\ \emph {et~al.}(2005)\citenamefont
  {Thiaville}, \citenamefont {Nakatani}, \citenamefont {Miltat},\ and\
  \citenamefont {Suzuki}}]{thiaville2005micromagnetic}%
  \BibitemOpen
  \bibfield  {author} {\bibinfo {author} {\bibfnamefont {A.}~\bibnamefont
  {Thiaville}}, \bibinfo {author} {\bibfnamefont {Y.}~\bibnamefont {Nakatani}},
  \bibinfo {author} {\bibfnamefont {J.}~\bibnamefont {Miltat}}, \ and\ \bibinfo
  {author} {\bibfnamefont {Y.}~\bibnamefont {Suzuki}},\ }\href@noop {}
  {\bibfield  {journal} {\bibinfo  {journal} {EPL}\ }\textbf {\bibinfo {volume}
  {69}},\ \bibinfo {pages} {990} (\bibinfo {year} {2005})}\BibitemShut
  {NoStop}%
\bibitem [{\citenamefont {Moore}\ \emph {et~al.}(2008)\citenamefont {Moore},
  \citenamefont {Miron}, \citenamefont {Gaudin}, \citenamefont {Serret},
  \citenamefont {Auffret}, \citenamefont {Rodmacq}, \citenamefont {Schuhl},
  \citenamefont {Pizzini}, \citenamefont {Vogel},\ and\ \citenamefont
  {Bonfim}}]{moore2008high}%
  \BibitemOpen
  \bibfield  {author} {\bibinfo {author} {\bibfnamefont {T.~A.}\ \bibnamefont
  {Moore}}, \bibinfo {author} {\bibfnamefont {I.}~\bibnamefont {Miron}},
  \bibinfo {author} {\bibfnamefont {G.}~\bibnamefont {Gaudin}}, \bibinfo
  {author} {\bibfnamefont {G.}~\bibnamefont {Serret}}, \bibinfo {author}
  {\bibfnamefont {S.}~\bibnamefont {Auffret}}, \bibinfo {author} {\bibfnamefont
  {B.}~\bibnamefont {Rodmacq}}, \bibinfo {author} {\bibfnamefont
  {A.}~\bibnamefont {Schuhl}}, \bibinfo {author} {\bibfnamefont
  {S.}~\bibnamefont {Pizzini}}, \bibinfo {author} {\bibfnamefont
  {J.}~\bibnamefont {Vogel}}, \ and\ \bibinfo {author} {\bibfnamefont
  {M.}~\bibnamefont {Bonfim}},\ }\href@noop {} {\bibfield  {journal} {\bibinfo
  {journal} {Appl. Phys. Lett.}\ }\textbf {\bibinfo {volume} {93}},\ \bibinfo
  {pages} {262504} (\bibinfo {year} {2008})}\BibitemShut {NoStop}%
\bibitem [{\citenamefont {Thiaville}\ and\ \citenamefont
  {Nakatani}(2006)}]{thiaville2006domain}%
  \BibitemOpen
  \bibfield  {author} {\bibinfo {author} {\bibfnamefont {A.}~\bibnamefont
  {Thiaville}}\ and\ \bibinfo {author} {\bibfnamefont {Y.}~\bibnamefont
  {Nakatani}},\ }in\ \href@noop {} {\emph {\bibinfo {booktitle} {Spin dynamics
  in confined magnetic structures III}}}\ (\bibinfo  {publisher} {Springer},\
  \bibinfo {year} {2006})\ pp.\ \bibinfo {pages} {161--205}\BibitemShut
  {NoStop}%
\bibitem [{\citenamefont {Mougin}\ \emph {et~al.}(2007)\citenamefont {Mougin},
  \citenamefont {Cormier}, \citenamefont {Adam}, \citenamefont {Metaxas},\ and\
  \citenamefont {Ferr{\'e}}}]{mougin2007domain}%
  \BibitemOpen
  \bibfield  {author} {\bibinfo {author} {\bibfnamefont {A.}~\bibnamefont
  {Mougin}}, \bibinfo {author} {\bibfnamefont {M.}~\bibnamefont {Cormier}},
  \bibinfo {author} {\bibfnamefont {J.}~\bibnamefont {Adam}}, \bibinfo {author}
  {\bibfnamefont {P.}~\bibnamefont {Metaxas}}, \ and\ \bibinfo {author}
  {\bibfnamefont {J.}~\bibnamefont {Ferr{\'e}}},\ }\href@noop {} {\bibfield
  {journal} {\bibinfo  {journal} {EPL}\ }\textbf {\bibinfo {volume} {78}},\
  \bibinfo {pages} {57007} (\bibinfo {year} {2007})}\BibitemShut {NoStop}%
\bibitem [{\citenamefont {Herranen}\ and\ \citenamefont
  {Laurson}(2015)}]{herranen2015domain}%
  \BibitemOpen
  \bibfield  {author} {\bibinfo {author} {\bibfnamefont {T.}~\bibnamefont
  {Herranen}}\ and\ \bibinfo {author} {\bibfnamefont {L.}~\bibnamefont
  {Laurson}},\ }\href@noop {} {\bibfield  {journal} {\bibinfo  {journal} {Phys.
  Rev. B}\ }\textbf {\bibinfo {volume} {92}},\ \bibinfo {pages} {100405}
  (\bibinfo {year} {2015})}\BibitemShut {NoStop}%
\bibitem [{\citenamefont {Herranen}\ and\ \citenamefont
  {Laurson}(2017)}]{herranen2017bloch}%
  \BibitemOpen
  \bibfield  {author} {\bibinfo {author} {\bibfnamefont {T.}~\bibnamefont
  {Herranen}}\ and\ \bibinfo {author} {\bibfnamefont {L.}~\bibnamefont
  {Laurson}},\ }\href@noop {} {\bibfield  {journal} {\bibinfo  {journal} {Phys.
  Rev. B}\ }\textbf {\bibinfo {volume} {96}},\ \bibinfo {pages} {144422}
  (\bibinfo {year} {2017})}\BibitemShut {NoStop}%
\bibitem [{\citenamefont {Thiaville}\ and\ \citenamefont
  {Miltat}(2018)}]{thiaville2018topology}%
  \BibitemOpen
  \bibfield  {author} {\bibinfo {author} {\bibfnamefont {A.}~\bibnamefont
  {Thiaville}}\ and\ \bibinfo {author} {\bibfnamefont {J.}~\bibnamefont
  {Miltat}},\ }in\ \href@noop {} {\emph {\bibinfo {booktitle} {Topology in
  Magnetism}}}\ (\bibinfo  {publisher} {Springer},\ \bibinfo {year} {2018})\
  pp.\ \bibinfo {pages} {41--73}\BibitemShut {NoStop}%
\bibitem [{\citenamefont {Garanin}\ \emph {et~al.}(2017)\citenamefont
  {Garanin}, \citenamefont {Chudnovsky},\ and\ \citenamefont
  {Zhang}}]{garanin2017skyrmion}%
  \BibitemOpen
  \bibfield  {author} {\bibinfo {author} {\bibfnamefont {D.~A.}\ \bibnamefont
  {Garanin}}, \bibinfo {author} {\bibfnamefont {E.~M.}\ \bibnamefont
  {Chudnovsky}}, \ and\ \bibinfo {author} {\bibfnamefont {X.}~\bibnamefont
  {Zhang}},\ }\href@noop {} {\bibfield  {journal} {\bibinfo  {journal} {EPL}\
  }\textbf {\bibinfo {volume} {120}},\ \bibinfo {pages} {17005} (\bibinfo
  {year} {2017})}\BibitemShut {NoStop}%
\bibitem [{\citenamefont {Malozemoff}\ and\ \citenamefont
  {Slonczewski}(1979)}]{malozemoff2016magnetic}%
  \BibitemOpen
  \bibfield  {author} {\bibinfo {author} {\bibfnamefont {A.}~\bibnamefont
  {Malozemoff}}\ and\ \bibinfo {author} {\bibfnamefont {J.}~\bibnamefont
  {Slonczewski}},\ }\href@noop {} {\emph {\bibinfo {title} {Magnetic Domain
  Walls in Bubble Materials: Advances in Materials and Device Research}}},\
  Vol.~\bibinfo {volume} {1}\ (\bibinfo  {publisher} {Academic press},\
  \bibinfo {year} {1979})\BibitemShut {NoStop}%
\bibitem [{\citenamefont {Konishi}(1983)}]{konishi1983new}%
  \BibitemOpen
  \bibfield  {author} {\bibinfo {author} {\bibfnamefont {S.}~\bibnamefont
  {Konishi}},\ }\href@noop {} {\bibfield  {journal} {\bibinfo  {journal} {IEEE
  Trans. Magn.}\ }\textbf {\bibinfo {volume} {19}},\ \bibinfo {pages} {1838}
  (\bibinfo {year} {1983})}\BibitemShut {NoStop}%
\bibitem [{\citenamefont {Vansteenkiste}\ \emph {et~al.}(2014)\citenamefont
  {Vansteenkiste}, \citenamefont {Leliaert}, \citenamefont {Dvornik},
  \citenamefont {Helsen}, \citenamefont {Garcia-Sanchez},\ and\ \citenamefont
  {Van~Waeyenberge}}]{vansteenkiste2014design}%
  \BibitemOpen
  \bibfield  {author} {\bibinfo {author} {\bibfnamefont {A.}~\bibnamefont
  {Vansteenkiste}}, \bibinfo {author} {\bibfnamefont {J.}~\bibnamefont
  {Leliaert}}, \bibinfo {author} {\bibfnamefont {M.}~\bibnamefont {Dvornik}},
  \bibinfo {author} {\bibfnamefont {M.}~\bibnamefont {Helsen}}, \bibinfo
  {author} {\bibfnamefont {F.}~\bibnamefont {Garcia-Sanchez}}, \ and\ \bibinfo
  {author} {\bibfnamefont {B.}~\bibnamefont {Van~Waeyenberge}},\ }\href@noop {}
  {\bibfield  {journal} {\bibinfo  {journal} {AIP Adv.}\ }\textbf {\bibinfo
  {volume} {4}},\ \bibinfo {pages} {107133} (\bibinfo {year}
  {2014})}\BibitemShut {NoStop}%
\bibitem [{\citenamefont {Weller}\ \emph {et~al.}(2000)\citenamefont {Weller},
  \citenamefont {Moser}, \citenamefont {Folks}, \citenamefont {Best},
  \citenamefont {Lee}, \citenamefont {Toney}, \citenamefont {Schwickert},
  \citenamefont {Thiele},\ and\ \citenamefont {Doerner}}]{weller2000high}%
  \BibitemOpen
  \bibfield  {author} {\bibinfo {author} {\bibfnamefont {D.}~\bibnamefont
  {Weller}}, \bibinfo {author} {\bibfnamefont {A.}~\bibnamefont {Moser}},
  \bibinfo {author} {\bibfnamefont {L.}~\bibnamefont {Folks}}, \bibinfo
  {author} {\bibfnamefont {M.~E.}\ \bibnamefont {Best}}, \bibinfo {author}
  {\bibfnamefont {W.}~\bibnamefont {Lee}}, \bibinfo {author} {\bibfnamefont
  {M.~F.}\ \bibnamefont {Toney}}, \bibinfo {author} {\bibfnamefont
  {M.}~\bibnamefont {Schwickert}}, \bibinfo {author} {\bibfnamefont {J.-U.}\
  \bibnamefont {Thiele}}, \ and\ \bibinfo {author} {\bibfnamefont {M.~F.}\
  \bibnamefont {Doerner}},\ }\href@noop {} {\bibfield  {journal} {\bibinfo
  {journal} {IEEE Trans. Magn.}\ }\textbf {\bibinfo {volume} {36}},\ \bibinfo
  {pages} {10} (\bibinfo {year} {2000})}\BibitemShut {NoStop}%
\bibitem [{\citenamefont {Martinez}(2011)}]{martinez2011stochastic}%
  \BibitemOpen
  \bibfield  {author} {\bibinfo {author} {\bibfnamefont {E.}~\bibnamefont
  {Martinez}},\ }\href@noop {} {\bibfield  {journal} {\bibinfo  {journal} {J.
  Phys. Condens. Matter}\ }\textbf {\bibinfo {volume} {24}},\ \bibinfo {pages}
  {024206} (\bibinfo {year} {2011})}\BibitemShut {NoStop}%
\bibitem [{SM()}]{SM}%
  \BibitemOpen
  \href@noop {} {}\bibinfo {note} {See Supplemental Material at [URL will be
  inserted by the publisher] for movies illustrating the DW dynamics
  responsible for the three different velocity drops.}\BibitemShut {Stop}%
\bibitem [{\citenamefont {Yoshimura}\ \emph {et~al.}(2016)\citenamefont
  {Yoshimura}, \citenamefont {Kim}, \citenamefont {Taniguchi}, \citenamefont
  {Tono}, \citenamefont {Ueda}, \citenamefont {Hiramatsu}, \citenamefont
  {Moriyama}, \citenamefont {Yamada}, \citenamefont {Nakatani},\ and\
  \citenamefont {Ono}}]{yoshimura2016soliton}%
  \BibitemOpen
  \bibfield  {author} {\bibinfo {author} {\bibfnamefont {Y.}~\bibnamefont
  {Yoshimura}}, \bibinfo {author} {\bibfnamefont {K.-J.}\ \bibnamefont {Kim}},
  \bibinfo {author} {\bibfnamefont {T.}~\bibnamefont {Taniguchi}}, \bibinfo
  {author} {\bibfnamefont {T.}~\bibnamefont {Tono}}, \bibinfo {author}
  {\bibfnamefont {K.}~\bibnamefont {Ueda}}, \bibinfo {author} {\bibfnamefont
  {R.}~\bibnamefont {Hiramatsu}}, \bibinfo {author} {\bibfnamefont
  {T.}~\bibnamefont {Moriyama}}, \bibinfo {author} {\bibfnamefont
  {K.}~\bibnamefont {Yamada}}, \bibinfo {author} {\bibfnamefont
  {Y.}~\bibnamefont {Nakatani}}, \ and\ \bibinfo {author} {\bibfnamefont
  {T.}~\bibnamefont {Ono}},\ }\href@noop {} {\bibfield  {journal} {\bibinfo
  {journal} {Nat. Phys.}\ }\textbf {\bibinfo {volume} {12}},\ \bibinfo {pages}
  {157} (\bibinfo {year} {2016})}\BibitemShut {NoStop}%
\bibitem [{\citenamefont {Boulle}\ \emph {et~al.}(2011)\citenamefont {Boulle},
  \citenamefont {Malinowski},\ and\ \citenamefont
  {Kl{\"a}ui}}]{boulle2011current}%
  \BibitemOpen
  \bibfield  {author} {\bibinfo {author} {\bibfnamefont {O.}~\bibnamefont
  {Boulle}}, \bibinfo {author} {\bibfnamefont {G.}~\bibnamefont {Malinowski}},
  \ and\ \bibinfo {author} {\bibfnamefont {M.}~\bibnamefont {Kl{\"a}ui}},\
  }\href@noop {} {\bibfield  {journal} {\bibinfo  {journal} {Mater. Sci. Eng. R
  Rep.}\ }\textbf {\bibinfo {volume} {72}},\ \bibinfo {pages} {159} (\bibinfo
  {year} {2011})}\BibitemShut {NoStop}%
\bibitem [{\citenamefont {Hubert}\ and\ \citenamefont
  {Sch{\"a}fer}(2008)}]{hubert2008magnetic}%
  \BibitemOpen
  \bibfield  {author} {\bibinfo {author} {\bibfnamefont {A.}~\bibnamefont
  {Hubert}}\ and\ \bibinfo {author} {\bibfnamefont {R.}~\bibnamefont
  {Sch{\"a}fer}},\ }\href@noop {} {\emph {\bibinfo {title} {Magnetic domains:
  the analysis of magnetic microstructures}}}\ (\bibinfo  {publisher} {Springer
  Science \& Business Media},\ \bibinfo {year} {2008})\BibitemShut {NoStop}%
\bibitem [{\citenamefont {Leliaert}\ \emph
  {et~al.}(2014{\natexlab{a}})\citenamefont {Leliaert}, \citenamefont {Van~de
  Wiele}, \citenamefont {Vansteenkiste}, \citenamefont {Laurson}, \citenamefont
  {Durin}, \citenamefont {Dupr{\'e}},\ and\ \citenamefont
  {Van~Waeyenberge}}]{leliaert2014numerical}%
  \BibitemOpen
  \bibfield  {author} {\bibinfo {author} {\bibfnamefont {J.}~\bibnamefont
  {Leliaert}}, \bibinfo {author} {\bibfnamefont {B.}~\bibnamefont {Van~de
  Wiele}}, \bibinfo {author} {\bibfnamefont {A.}~\bibnamefont {Vansteenkiste}},
  \bibinfo {author} {\bibfnamefont {L.}~\bibnamefont {Laurson}}, \bibinfo
  {author} {\bibfnamefont {G.}~\bibnamefont {Durin}}, \bibinfo {author}
  {\bibfnamefont {L.}~\bibnamefont {Dupr{\'e}}}, \ and\ \bibinfo {author}
  {\bibfnamefont {B.}~\bibnamefont {Van~Waeyenberge}},\ }\href@noop {}
  {\bibfield  {journal} {\bibinfo  {journal} {J. Appl. Phys.}\ }\textbf
  {\bibinfo {volume} {115}},\ \bibinfo {pages} {17D102} (\bibinfo {year}
  {2014}{\natexlab{a}})}\BibitemShut {NoStop}%
\bibitem [{\citenamefont {Leliaert}\ \emph
  {et~al.}(2014{\natexlab{b}})\citenamefont {Leliaert}, \citenamefont {Van~de
  Wiele}, \citenamefont {Vansteenkiste}, \citenamefont {Laurson}, \citenamefont
  {Durin}, \citenamefont {Dupr{\'e}},\ and\ \citenamefont
  {Van~Waeyenberge}}]{leliaert2014current}%
  \BibitemOpen
  \bibfield  {author} {\bibinfo {author} {\bibfnamefont {J.}~\bibnamefont
  {Leliaert}}, \bibinfo {author} {\bibfnamefont {B.}~\bibnamefont {Van~de
  Wiele}}, \bibinfo {author} {\bibfnamefont {A.}~\bibnamefont {Vansteenkiste}},
  \bibinfo {author} {\bibfnamefont {L.}~\bibnamefont {Laurson}}, \bibinfo
  {author} {\bibfnamefont {G.}~\bibnamefont {Durin}}, \bibinfo {author}
  {\bibfnamefont {L.}~\bibnamefont {Dupr{\'e}}}, \ and\ \bibinfo {author}
  {\bibfnamefont {B.}~\bibnamefont {Van~Waeyenberge}},\ }\href@noop {}
  {\bibfield  {journal} {\bibinfo  {journal} {J. Appl. Phys.}\ }\textbf
  {\bibinfo {volume} {115}},\ \bibinfo {pages} {233903} (\bibinfo {year}
  {2014}{\natexlab{b}})}\BibitemShut {NoStop}%
\bibitem [{\citenamefont {Thiaville}\ \emph {et~al.}(2012)\citenamefont
  {Thiaville}, \citenamefont {Rohart}, \citenamefont {Ju{\'e}}, \citenamefont
  {Cros},\ and\ \citenamefont {Fert}}]{thiaville2012dynamics}%
  \BibitemOpen
  \bibfield  {author} {\bibinfo {author} {\bibfnamefont {A.}~\bibnamefont
  {Thiaville}}, \bibinfo {author} {\bibfnamefont {S.}~\bibnamefont {Rohart}},
  \bibinfo {author} {\bibfnamefont {{\'E}.}~\bibnamefont {Ju{\'e}}}, \bibinfo
  {author} {\bibfnamefont {V.}~\bibnamefont {Cros}}, \ and\ \bibinfo {author}
  {\bibfnamefont {A.}~\bibnamefont {Fert}},\ }\href@noop {} {\bibfield
  {journal} {\bibinfo  {journal} {EPL}\ }\textbf {\bibinfo {volume} {100}},\
  \bibinfo {pages} {57002} (\bibinfo {year} {2012})}\BibitemShut {NoStop}%
\end{thebibliography}%

\end{document}